\documentclass[usenatbib,useAMS,preprint,psfig2]{mn2e}
\usepackage{psfig,graphicx}

\def\gapprox{\lower.4ex\hbox{$\;\buildrel >\over{\scriptstyle\sim}\;$}}
\def\lapprox{\lower.4ex\hbox{$\;\buildrel <\over{\scriptstyle\sim}\;$}}

\def\bk{\mbox{\boldmath $k$}}
\def\bhk{\hat{\mbox{\boldmath $k$}}}
\def\bv{\mbox{\boldmath $v$}}
\def\bx{\mbox{\boldmath $x$}}
\def\be{\mbox{\boldmath $e$}}
\def\bj{\mbox{\boldmath $j$}}
\def\bn{\mbox{\boldmath $n$}}
\def\bB{\mbox{\boldmath $B$}}

\title{Coherent synchrotron emission from cosmic ray air showers}
                                                                                                            
\author[Luo]
      {Qinghuan Luo\\
        School of Physics, The University of Sydney, NSW 2006, Australia\\
}
                                                                                                            
\date{
          --- Received
         in original form February, 2006
        }
                                                                                                            
\pubyear{2006}
                                                                                                            
\begin{document}
                                                                                                            
\maketitle
                                                                                                            
\begin{abstract}
Coherent synchrotron emission by particles moving along semi-infinite tracks
is discussed, with a specific application to radio emission from air showers 
induced by high-energy cosmic rays.  It is shown that in general, radiation from a particle
moving along a semi-infinite orbit consists of usual synchrotron emission
and modified impulsive bremsstrahlung. The latter component is due to the instantaneous
onset of the curved trajectory of the emitting particle at its creation.
Inclusion of the bremsstrahlung leads to broadening of the radiation pattern and a 
slower decay of the spectrum
at the cut-off frequency than the conventional synchrotron emission. 
Possible implications of these features for air shower radio emission are discussed.
\end{abstract}


\section{Introduction}

There is recently growing attention on radio detection of high energy
cosmic rays~\citep{dz89,zhs92,heo96,avz00,fg03,g-etal04,fgp04,f-etal05}.
High energy cosmic rays can initiate an electromagnetic 
cascade in a medium where relativistic electrons and positrons can be produced in a
volume with a longitudinal (along the line of sight) dimension being smaller 
than the relevant radio wavelength. So, particles form a coherent bunch, acting 
like a single charged particle that emits a short burst of coherent radio emission.
Two radiation processes have been considered: coherent Cerenkov radiation 
\citep{a62, a65} and coherent synchrotron radiation~\citep{fg03,hf03}. 
(Coherent radio emission from air showers was first considered by \cite{kl66}, also
\cite{c67}, but their theory was not explicitly based on geosynchrotron
emission.) The former requires charge asymmetry, say an excess of electrons, and
a relatively dense medium for coherent Cerenkov emission to be at radio frequency.
For example, the Moon is a good target for high energy neutrinos that can lead to a
cascade in the lunar rocks. Excess electrons can develop leading to coherent Cerenkov
emission at radio frequency with wavelength comparable with or larger than the longitudinal
size of the cascade~\citep{a62,a65}. For air showers, coherent 
synchrotron emission is generally considered more important
than Cerenkov radiation (in the radio band)~\citep{fg03,g-etal04}. 
The shower produces a bunch of relativistic electrons and positrons 
emitting coherent  synchrotron radiation in geomagnetic fields.
Unlike coherent Cerenkov radiation, coherent synchrotron emission does not need
charge asymmetry~\citep{fg03}.

So far, the relevant spectra of coherent synchrotron emission from air showers 
were commonly calculated using numerical simulation based on the retarded-potential 
method \citep{sgr03,hf05a,hf05b}. In this method, the radiation is calculated from
the retarded potential~\citep{j98}. Although coherent synchrotron emission has been 
considered analytically \citep{ab02,fg03,hf03}, their 
calculation is based on the standard synchrotron radiation formula, which does not include the 
effect due to the particle's finite track. For air showers, effective coherent emission occurs
in the core of the shower where most of radiating particles are created. Thus, it is of interest
to consider radiation by a particle moving along a semi-infinite trajectory.
Apart from the usual synchrotron emission there is emission due to the onset of 
the particle's curved trajectory. The latter component is referred to
as the modified impulsive bremsstrahlung (MIB) as it is due to the combined
effect of the usual impulsive (or prompt) bremsstrahlung due to particle ($e^\pm$) creation,
which is modeled as an abrupt jump in the particle's velocity from zero to $c$,
and curvature of the particle's trajectory. It is worth noting that the finite 
track effect was considered for Cerenkov radiation~\citep{t39}  and was taken 
into account explicitly in calculation of cosmic-ray induced showers 
in a dense medium~\citep{zhs92}. 
In the case of Cerenkov radiation, the finite track leads to a reduction in radiation 
intensity and modification of angular distribution of the emission. 
Since the main objective of studying radio
emission from air showers is to infer the properties of the cosmic rays that
induce the showers, one needs to determine the radio spectrum accurately.
In this paper, we present a general formalism for coherent synchrotron 
radiation from a nonstationary many-particle system, which takes account 
of MIB. The formalism developed here is based on the 
single particle treatment~\citep{mm91}, in which radiation is due to a current associated with 
particle's motion in a medium. Here, the current is regarded as extraneous as it
is different from that due to the plasma response to waves. The spectrum of
radiation from a many-particle system is derived from the total current that is obtained 
by adding all the currents due to individual particles. In the relativistic limit 
as in the usual synchrotron radiation, the spectrum can be expressed in terms of 
the Airy functions and as a result, the radiation is highly beamed.

In Sec.2, a general formalism for synchrotron emission from a many-particle system
is derived by including the effect of MIB due to
the effect of a particle's semi-infinite track. Coherent synchrotron emission is
considered in Sec. 3 and the application to air showers is discussed in Sec. 4.

\section{Energy spectrum}

In cosmic ray air showers, electrons and positrons are created with a relativistic 
velocity, emitting synchrotron radiation in the geomagnetic field. Effective coherent 
emission by these secondary particles occurs at the core of the shower located very close to
where most particles are produced. Thus, the finite track effect, in particular
the initial position of the particle's orbit, can be important and needs to be included
explicitly in the calculation of the synchrotron spectral power. In the following, we 
start with the single particle formalism.
 
\subsection{Single particle treatment}

Consider a charged ($q$) particle created at $t=0$ moving 
along a trajectory $\bx(t)$ with a flight time $T$. 
The current associated with the particle is
\begin{equation}
{\bj}(\omega,{\bk})=q\int^{T}_{0}dt\,{\bv}(t)\exp\left[i\left(
\omega t-{\bk}\cdot {\bx}(t)\right)\right],
\label{eq:c1}
\end{equation} 
where ${\bv}=d{\bx}/dt$ is the particle's velocity. The current 
(\ref{eq:c1}) can be regarded as an extraneous current due to a single 
particle's motion (as compared to the induced current due to plasma response). 
In the usual application to radiation in astrophysical plasmas, 
the time integration is taken from $-\infty$ to $\infty$ ~\citep{m86}.
A finite $T$, particularly the initial point at $t=0$, introduces a 
boundary (finite track) effect into (\ref{eq:c1}). The energy spectrum 
can be found from the expression \citep{mm91}
\begin{equation}
U_M({\bk})={1\over2\varepsilon_0}|{\be}^*_M\cdot{\bj}|^2, 
\label{eq:sp1}
\end{equation}
where ${\be}^*_M$ is the complex conjugate of the polarization $\be_M$ of the 
wave emitted in the mode $M$. The single particle's spectral power can be 
derived from $P_M=U_M/T$.

The orbit of a charged particle spiraling in a magnetic field $\bB$ oriented 
along the $z$-axis can be written as
\begin{eqnarray}
{\bx}(t)&=&{\bx}_{0c}+{v_{\perp }\over \Omega}
\biggl[\sin\bigl(\Omega t-\psi_{0}\bigr){\be}_x+
\cos\bigl(\Omega t-\psi_{0}\bigr){\be}_y\biggr]
\nonumber\\
&&
+v_{\parallel } t\,{\be}_z,
\end{eqnarray}
where $t\geq 0$, ${\bx}_{0c}$ is the initial position of the particle's gyrocenter, $\psi_{0}$ 
is the initial gyrophase, which is defined here as the the azimuthal angle of the particle's initial
velocity relative to the magnetic field, $\Omega=\eta\Omega_e/\gamma$,
$q=\eta e$, $\eta$ is the charge sign,
$\gamma=1/(1-v^2/c^2)^{1/2}$ is the Lorentz factor, and $\Omega_e=eB/m_e$ is 
the gyrofrequency. The standard method to calculate the current is to expand the 
exponential term in terms of Bessel functions~\citep{mm91}. In the relativistic 
limit as in the case relevant here, instant emission can only be seen during a very short
time interval $\Delta t\sim R_g/c\gamma^3=\beta_\perp/\Omega_e\gamma^2$,
where $R_g=v_\perp/\Omega$ is the gyroradius. Thus, the orbit can be expanded 
on $t\Omega_e\sim \beta_\perp/\gamma^2\ll1$ and the exponential in (\ref{eq:c1}) can
be expressed into the form
\begin{equation}
\omega t-{\bk}\cdot{\bx}\approx d+a\tau+b\tau^3,
\end{equation}
where $\tau=t+t_0$ with $t_0=\Omega^{-1}\tan(\phi-\psi_0)$ and
\begin{eqnarray}
a&=&\omega-k_\parallel v_\parallel\nonumber\\
&&-k_\perp v_\perp\biggl[
\cos(\phi-\psi_0)+\textstyle{1\over2}
\tan(\phi-\psi_0)\sin(\phi-\psi_0)\biggr],
\nonumber\\
b&=&\textstyle{1\over6}
k_\perp v_\perp \Omega^2\,\cos(\phi-\psi_0),\nonumber\\
d&=&-{\bk}\cdot{\bx}_{0c}-{1\over\Omega}
\biggl[
\omega-k_\parallel v_\parallel\nonumber\\
&&-\textstyle{2\over3}k_\perp v_\perp
{\displaystyle{\sin^2(\phi-\psi_0)}\over\displaystyle{
\cos(\phi-\psi_0)}}\biggr]
\tan(\phi-\psi_0).
\label{eq:d}
\end{eqnarray}
We assume the observation direction is $\bhk={\bk}/k=(\sin\theta\cos\phi,
\sin\theta\sin\phi,\cos\theta)$ and define
spherical coordinates ${\bhk}={\be}_r$.
The projection of the current to the plane perpendicular to ${\bk}$ can be written as
${\bj}_\perp\equiv{\bj}-{\bhk}{\bhk}\cdot{\bj}$, which has the
following components:
\begin{eqnarray}
j_{\perp\theta}&\approx&
e\eta\Biggl[
v_{\perp }\biggl({F\over\cos(\phi-\psi_{0})}\nonumber\\
&&+i\Omega F'\sin(\phi-\psi_{0})
\biggr)\cos\theta- v_{\parallel }F\sin\theta\Biggr]\,e^{id},\nonumber\\
j_{\perp\phi}&\approx& ie\eta v_{\perp }\Omega F'\cos(\phi-\psi_{0})\,e^{id},
\label{eq:c3}
\end{eqnarray}
where the relevant integrals are defined as
\begin{eqnarray}
F&=&\int^{T+t_0}_{t_0}d\tau\,e^{i(a\tau+b{\tau}^3)},\nonumber\\
F'&\equiv& {\partial F\over \partial a}=
i\int^{T+t_0}_{t_0}d\tau\,\tau\,e^{i(a\tau+b{\tau}^3)}.
\label{eq:FF1}
\end{eqnarray}
The relativistic beaming ($\gamma\gg1$) 
implies that $|\phi-\psi_0|\ll1$ and $|\alpha-\theta|\ll1$,
where $\alpha=\arctan(v_\perp/v_\parallel)$ is the pitch angle.
In these approximations, one has $a\approx \omega[1-n\beta\cos(\theta-\alpha)]\approx
(\omega/2)[2(1-n)+\gamma^{-2}+(\theta-\alpha)^2]$ and $b\approx
(n\omega\Omega^2/6)\cos(\phi-\psi_0)\sin\theta\sin\alpha$,
where the refraction index is assumed to satisfy $|1-n|\ll1$.

\subsection{Spectrum}

\begin{figure}
\includegraphics[width=7cm]{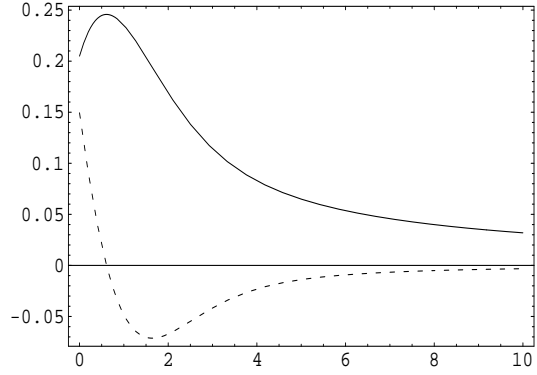}
\caption{Plots of ${\rm Gi}(\xi)$ (solid) and ${\rm Gi}'(\xi)$ (dashed).
}
\label{fig:Gi}
\end{figure}

One may write the energy spectrum as the energy radiated per unit frequency per
unit solid angle, $U(\omega,\hat{\bk})=(\omega^2/8\pi^3c^3)\sum_MU_M({\bk})$,
where $U_M({\bk})$ is given by (\ref{eq:sp1}) and the summation is made over
two orthogonal modes. One finds
\begin{eqnarray}
U(\omega,\bhk)&=&{\omega^2\over16\pi^3\varepsilon_0c^3}
\Bigl(|j_{\perp\theta}|^2+
|j_{\perp\phi}|^2\Bigr)\nonumber\\
&\approx&
{e^2\omega^2\over16\pi^3\varepsilon_0c}\biggl[
(\alpha-\theta)^2|F|^2 +
{\Omega^2_e\over\gamma^2}\,
\sin^2\alpha\,|F'|^2\biggr].
\label{eq:P2}
\end{eqnarray}
The integrals in (\ref{eq:FF1}) can be expressed in terms of the Airy functions
provided that the flight time $T$ is much longer than the 
duration of the synchrotron pulse ($\Delta t\sim 1/\Omega_e\gamma^2$).
In this approximation, one may take the limit $T+t_0\to \infty$ and the integrals
reduce to the form
\begin{equation}
F\approx{\pi\over (3b)^{1/3}}\Biggl[\,{\rm Ai}(\xi)+
i{\rm Gi}(\xi)\Biggr]-\Phi(a,b),
\label{eq:F2}
\end{equation}
\begin{equation}
F'\approx{\pi\over (3b)^{2/3}}\Biggl[\,{\rm Ai}'\left(\xi\right)+
i{\rm Gi}'(\xi)\Biggr]-{\partial\Phi(a,b)\over\partial a},
\label{eq:Fp2}
\end{equation}
\begin{equation}
\xi={a\over(3b)^{1/3}}\approx \left({\omega\over2\Omega}\right)^{2/3}
\,{2(1-n)+\gamma^{-2}+(\theta-\alpha)^2\over
[\cos(\phi-\psi_0)\,\sin\theta\,\sin\alpha]^{1/3}},
\end{equation}
\begin{equation}
\Phi(a,b)=\int^{t_0}_0d\tau\,e^{i(a\tau+b\tau^3)},
\label{eq:Phi}
\end{equation}
where ${\rm Gi}(\xi)=(1/3){\rm Bi}(\xi)-
\int^\xi_0[{\rm Ai}(\xi'){\rm Bi}(\xi)-{\rm Ai}(\xi){\rm Bi}(\xi')]d\xi'$,
${\rm Ai}(\xi)$ and ${\rm Bi}(\xi)$ are the Airy functions (see Abramowitz \& Stegun 1970).
Similar to ${\rm Ai}(\xi)$ and $-{\rm Ai}'(\xi)$, both ${\rm Gi}(\xi)$ and $-{\rm Gi}'(\xi)$ 
are a decaying function for $\xi\geq0$ (as shown in figure~\ref{fig:Gi}). 
The first terms in both square brackets on the right-hand sides of 
(\ref{eq:F2}) and (\ref{eq:Fp2}) correspond to synchrotron radiation 
for a particle moving along a semi-infinite trajectory.  
The terms ${\rm Gi}$ and ${\rm Gi}'$ describe MIB arising from the combined 
effect of the semi-infinite track's initial point and curvature.
One can show that MIB has features of usual impulsive bremsstrahlung
(cf. Eq \ref{eq:Ub2}),
i.e. emission due to an instantaneous change in the velocity from zero 
to $v\sim c$ at $t=0$ (cf. Eq. 1) \citep{ll71,g03}. One should emphasize
here that the impulsive bremsstrahlung considered here is different from 
the usual bremsstrahlung by a charged particle interacting with 
the Coulomb field of nuclei in matter. It can be shown that the last terms 
($\Phi$ and $\partial\Phi/\partial a$)
on the right-hand side of both (\ref{eq:F2}) and (\ref{eq:Fp2}) can be 
ignored provided that $|\phi-\psi_0|\gamma\ll\pi(2\Omega_e\gamma^2/\omega)^{1/3}$. 
The characteristic frequency can be estimated from $a^3\sim b$, which leads to
$\omega\sim \gamma^2\Omega_e$. 

The spectrum can be written as sum of usual synchrotron emission 
($U_{\rm syn}$) from a semi-finite track and MIB emission ($U_{\rm b}$) 
due to the velocity jump at $t=0$ and the trajectory's curvature, that is,
\begin{equation}
U=U_{\rm syn}+U_{\rm b},
\label{eq:U}
\end{equation}
with
\begin{eqnarray}
U_{\rm syn}
&\approx&{e^2\over16\pi^3\varepsilon_0c}\left(
{\omega\over a}\right)^2\Biggl[
(\theta-\alpha)^2\Bigl(\pi\xi{\rm Ai}(\xi)\Bigr)^2\nonumber\\
&&+
\left({\Omega_e\sin\alpha\over a\gamma }\right)^2
\Bigl(\pi\xi^2{\rm Ai}'(\xi)\Bigr)^2
\Biggr],
\label{eq:Usyn}
\end{eqnarray}
\begin{eqnarray}
U_{\rm b}
&\approx&{e^2\over16\pi^3\varepsilon_0c}\left(
{\omega\over a}\right)^2\Biggl[
(\theta-\alpha)^2\Bigl(\pi\xi{\rm Gi}(\xi)\Bigr)^2\nonumber\\
&&+
\left({\Omega_e\sin\alpha\over a\gamma }\right)^2
\Bigl(\pi\xi^2{\rm Gi}'(\xi)\Bigr)^2
\Biggr],
\label{eq:Ub}
\end{eqnarray}
where we neglect both the term (\ref{eq:Phi}) and its derivative. 
It is often convenient to separate the radiation into
two components, with one polarized perpendicular to the orbit plane,
corresponding to the first terms in (\ref{eq:Usyn}) and (\ref{eq:Ub}), and 
the other polarized in the plane, corresponding to the second terms involving derivatives.
Specifically, one may write $U_{\rm syn}=U_{\rm syn \perp}+U_{\rm syn \parallel}$ and
$U_{\rm b}=U_{\rm b \perp}+U_{\rm b \parallel}$. 
Notice that $U_{\rm syn}$ is smaller by a factor of 4
than that for normal synchrotron emission from a full infinite track (cf. Eq. ~\ref{eq:syn}).
Taking the limit $\Omega_e\to 0$ one can easily verify that $U_{\rm b}$ reduces to the familar form for
the prompt bremsstrahlung in the case where a particle's velocity abruptly changes from 
zero to a constant velocity~\citep{ll71}. In this limit, the angle $\alpha$ is re-interpreted as an
angle that the velocity makes with respect to the $z$ axis.  Since one has $\xi\to \infty$ and hence
$\pi\xi{\rm Gi}(\xi)\to 1$ (Abramowitz \& Stegun 1970), one finds
\begin{equation}
U_{\rm b}(\omega,\bhk)\approx
{e^2\over4\pi^3\varepsilon_0c}\left[{\theta-\alpha\over
\gamma^{-2}+(\theta-\alpha)^2}\right]^2,
\label{eq:Ub2}
\end{equation}
which is the same as that for the case where a particle instaneously acquires a constant 
velocity~\citep{ll71}.

Apart from its property of the usual impulsive bremsstrahlung, MIB also has a 
feature of synchrotron radiation (cf. Sec. 2.3), i.e. $U_b$ increases with frequency
as a power-law, similar to the usual synchrotron spectrum at low frequencies. 
Such similarity can be understood as due to that the emitting particle's 
trajectory is curved, while for the usual impulsive bresstrahlung
the particle's trajectory is a straight line. Because of this similar
feature to synchrotron radiation MIB is nonzero when one applies it to pair 
creation. 

\subsection{Usual synchrotron radiation}

The conventional synchrotron emission can be reproduced by adding the other 
semi-infinite orbit from $-\infty$ to 0 in (\ref{eq:c1}). 
Specifically, one first drops the terms $\Phi$ in (\ref{eq:F2}) and $\partial\Phi/\partial a$
in (\ref{eq:Fp2}), both of which are cancelled out by the corresponding terms 
from the negative semi-infinite trajectory, and then 
replaces $F$ and $F'$ in (\ref{eq:P2}) with $2{\rm Re}(F)$ and $2{\rm Re}(F')$, respectively.
Thus, the bremsstrahlung terms cancel out.
One then reproduces the usual synchrotron formula~\citep{mm91}
\begin{eqnarray}
U(\omega,\bhk)&=&4U_{\rm syn}\nonumber\\
&\approx&{e^2\over6\pi^3\varepsilon_0c}\left(
{\omega\over\Omega\sin\alpha}\right)^2
\biggl[\gamma^{-2}+(\theta-\alpha)^2\biggr]^2
\nonumber\\
&&\times
\Biggl[
{(\theta-\alpha)^2\over\gamma^{-2}+(\theta-\alpha)^2}
{\rm K}^2_{1/3}(\rho)+{\rm K}^2_{2/3}(\rho)\Biggr],
\label{eq:syn}
\end{eqnarray}
where $n=1$ and $\rho=(2/3)\xi^{3/2}$. We rewrite the Airy functions in terms of 
the modified Bessel functions and the approximations $|\theta-\alpha|\ll1$ and
$|\phi-\psi_0|\ll1$ are used. 

\begin{figure}
\includegraphics[width=6.8cm]{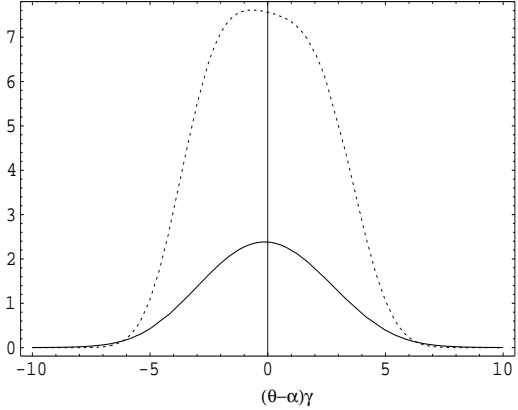}
\caption{Plot $U_\Vert$ (arbitrary scale) as a function of
$(\theta-\alpha)\gamma$ with $\phi=\psi_0$ for a single particle.
The dashed line corresponds to the usual synchrotron emission with 
$U_\parallel=4U_{{\rm syn}\parallel}$, in the absence of MIB
emission. We assume $\gamma=80$, $n=1$, the gyrofrequency $\Omega_e=5\times10^6\,{\rm
s}^{-1}$, $\omega/2\pi=100\,\rm MHz$ and the pitch angle $\alpha=\pi/4$.
}
\label{fig:profile1}
\end{figure}

\begin{figure}
\includegraphics[width=7cm]{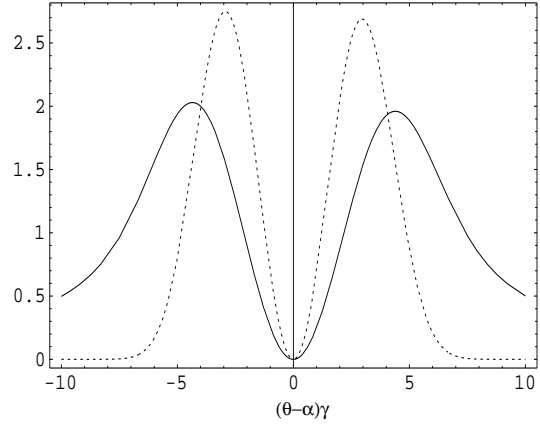}
\caption{Plot $U_\perp$ (arbitrary scale) as a function of
$(\theta-\alpha)\gamma$ with $\phi=\psi_0$. The parameters are as in figure~\ref{fig:profile1}.
Apart from a reduction in intensity, the modified synchrotron emission (usual
synchrotron plus MIB) has a much
wider profile than the usual synchrotron emission.
}
\label{fig:profile2}
\end{figure}

\begin{figure}
\includegraphics[width=7cm]{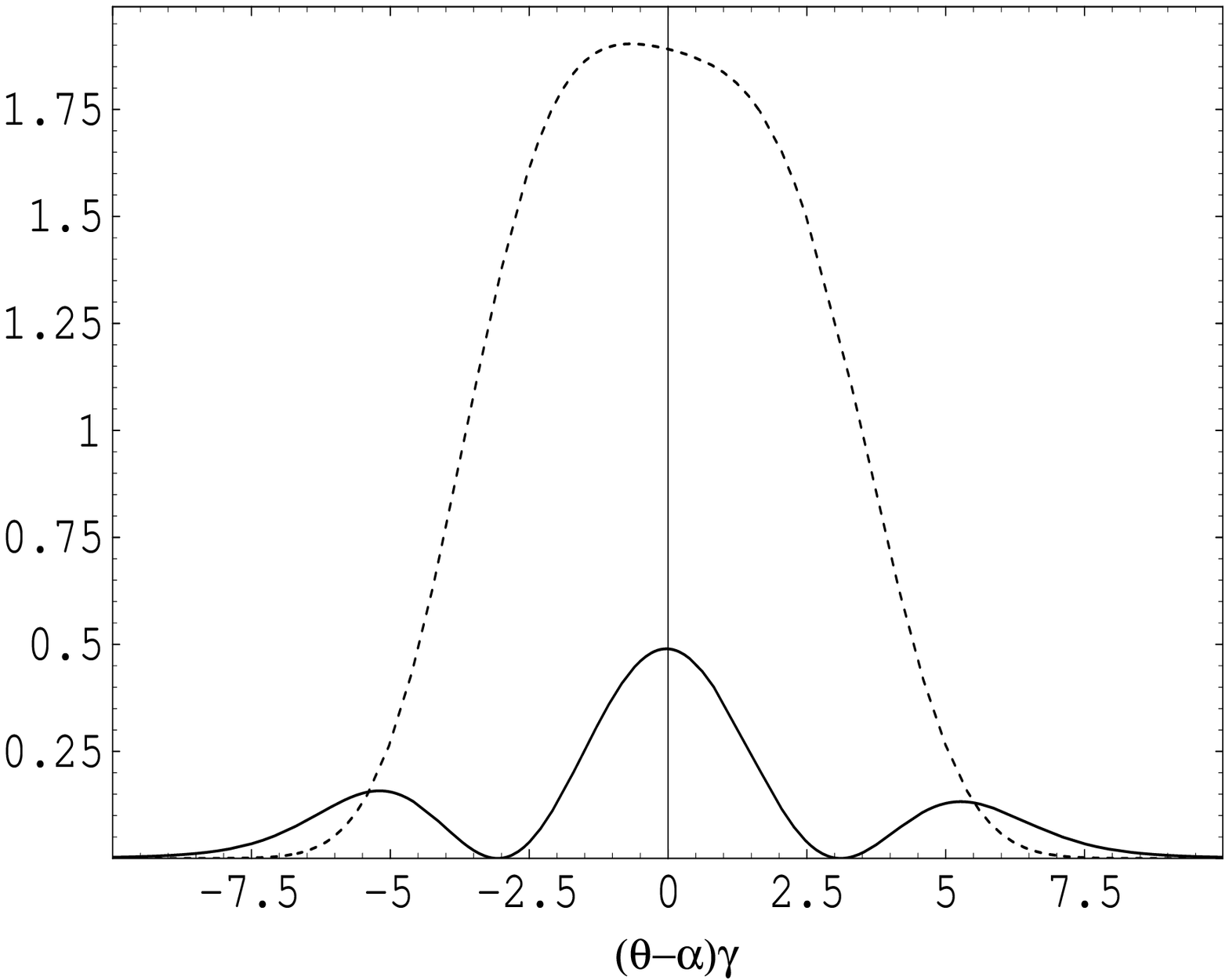}
\caption{Angular profile (arbitrary scale) as in figure~\ref{fig:profile1}.
The solid and dashed lines correspond respectively to 
$U_{{\rm b}\parallel}$ and $U_{{\rm syn}\parallel}$.
}
\label{fig:profile3}
\end{figure}

\begin{figure}
\includegraphics[width=7cm]{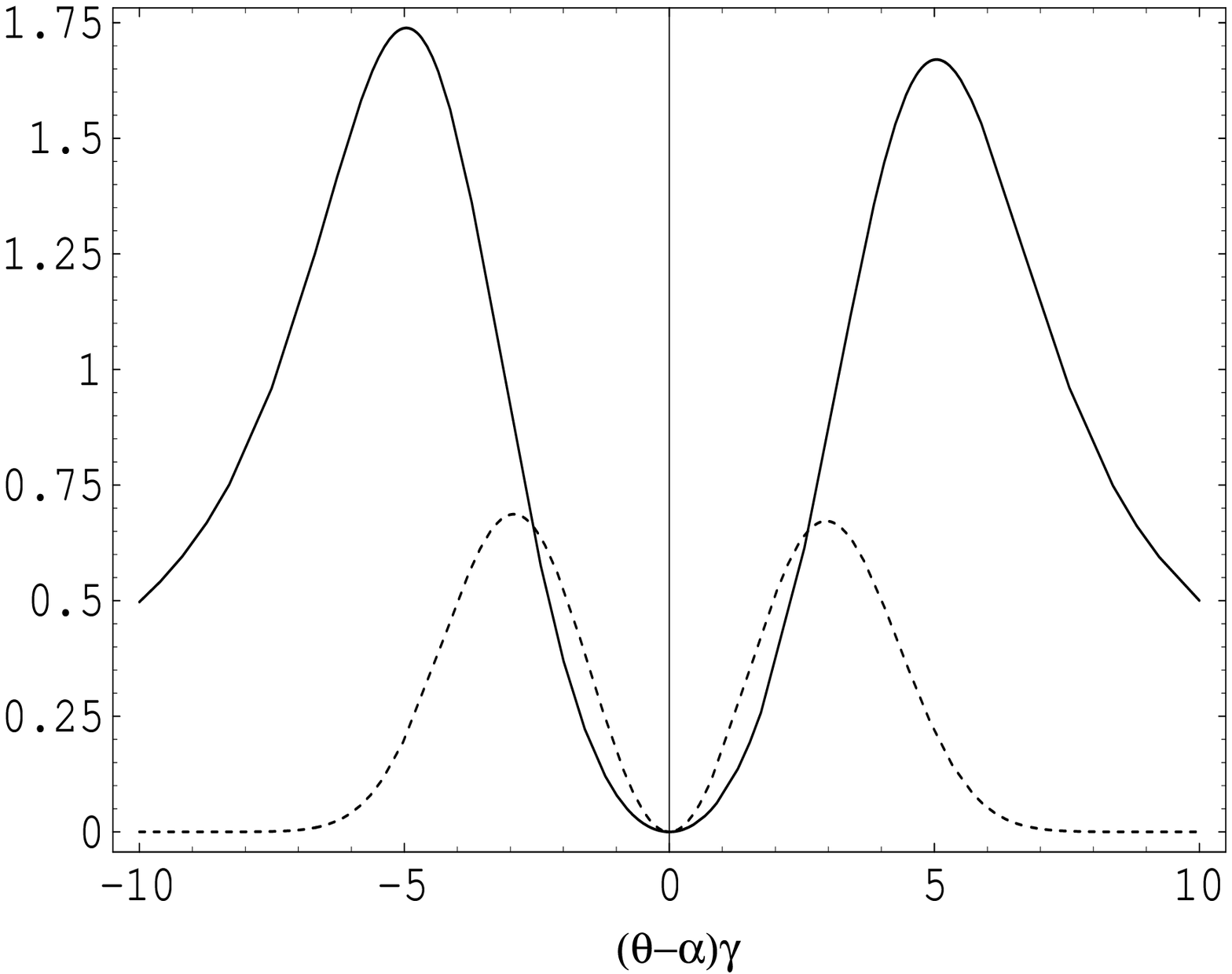}
\caption{Angular profile (arbitrary scale) as in figure~\ref{fig:profile1}.
The solid and dashed lines correspond respectively to 
$U_{{\rm b}\perp}$ and $U_{{\rm syn}\perp}$.
}
\label{fig:profile4}
\end{figure}

\begin{figure}
\includegraphics[width=7.5cm]{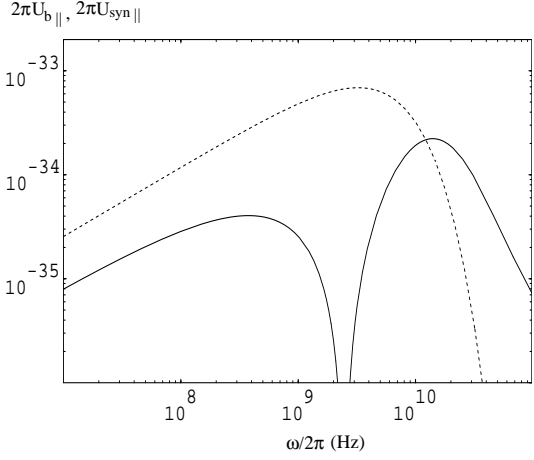}
\caption{Energy spectrum (in ${\rm J}\,{\rm Hz}^{-1}\,{\rm sr}^{-1}$)
as a function of frequency $\omega/2\pi$ (Hz) for
$\phi=\psi_0$ for a single particle. The solid and dashed lines represent 
$2\pi U_{{\rm b}\parallel}$ and $2\pi U_{{\rm syn}\parallel}$, respectively.
We assume $(\theta-\alpha)\gamma=0.5$, $N=10^7$, $\gamma=80$,
$n=1$, $\Omega_e=5\times10^6\,{\rm s}^{-1}$, $(\theta-\alpha)\gamma=0.5$,
and $\psi=\psi_0$.  At low frequencies, the spectrum is very similar 
to that for the usual synchrotron, i.e. $U\propto\omega^{2/3}$ for $\omega\ll
\omega_c\sim\Omega_e\gamma^2$. 
}
\label{fig:spectrum1}
\end{figure}

\begin{figure}
\includegraphics[width=7.5cm]{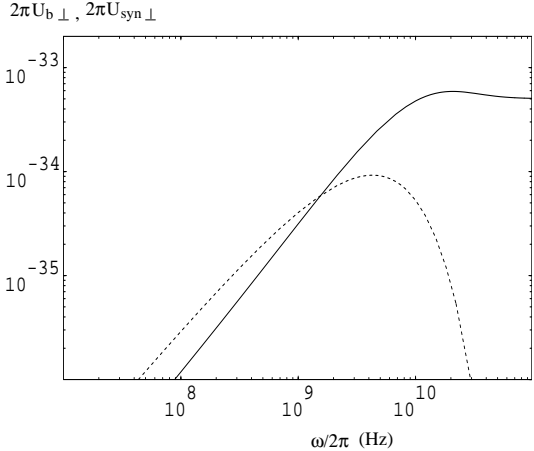}
\caption{Energy spectrum as in figure~\ref{fig:spectrum1} for the
$\perp$-polarized components. Notice that the spectral component $2\pi U_{{\rm b}\perp}$
(solid line) tends to level out at high frequencies where the approximate
form (\ref{eq:Ub2}) applies.
}
\label{fig:spectrum2}
\end{figure}

\begin{figure}
\includegraphics[width=7.5cm]{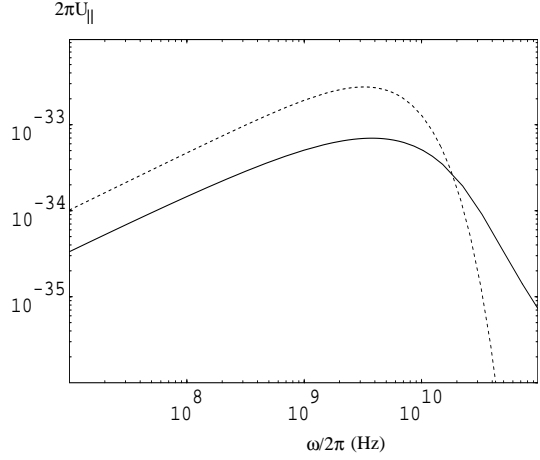}
\caption{Energy spectrum $2\pi U_\Vert=2\pi(U_{{\rm b}\parallel}+U_{{\rm syn}\parallel})$ 
(in ${\rm J}\,{\rm Hz}^{-1}\,{\rm sr}^{-1}$).
The dashed line corresponds to synchrotron emission $8\pi U_{{\rm syn}\parallel}$. 
The parameters are as in figure~\ref{fig:spectrum1}.
In comparison with the normal synchrotron emission
(dashed line), due to the MIB the spectrum (solid line) drops
off much slower than the exponential decay at $\omega\gg\omega_c$.
}
\label{fig:spectrum3}
\end{figure}

The angular distribution of single particle's spectrum (\ref{eq:U}) is shown in
figures~\ref{fig:profile1}-\ref{fig:profile4}. Here, the radiation is
separated into the parallel component, $U_\parallel=U_{\rm syn\parallel}+
U_{\rm b \parallel}$ and perpendicular component, $U_\perp=
U_{\rm syn\perp}+U_{\rm b \perp}$.
The inclusion of the boundary effect leads to an overall reduction in intensity
and a significant broadening of the angular profile.
Figures~\ref{fig:spectrum1} and \ref{fig:spectrum2} show a comparison 
of the two components $U_{\rm b}$ and $U_{\rm syn}$ as a function of frequency.
These two components are similar to each other at low frequencies. However, 
$U_{{\rm b}\parallel}$ has two cutoffs, with the lower one determined from 
the zero of ${\rm Gi}'(\xi)$ (cf. figure~\ref{fig:Gi}). For the perpendicular polarization, 
$U_{{\rm b}\perp}$ levels out at high frequencies and behaves much like the usual 
impulsive bremsstrahlung.
The energy spectrum for the parallel polarization is shown in figure~\ref{fig:spectrum3}.
The synchrotron emission drops off exponentially above the critical frequency.
Since ${\rm Gi}(\xi)\sim 1/\xi$ and ${\rm Gi}'(\xi)\sim1/\xi^2$, which drop
off much slower than the exponential decay of ${\rm Ai}(\xi)$ and ${\rm Ai}'(\xi)$ 
at a large $\xi$, the bremsstrahlung component decays much slower than the 
usual synchrotron emission.
 
\subsection{Many-particle system}

The single particle formula can be extended to a many-particle system
by adding all the currents from individual particles. Let the $s$th particle
be created at time $t_s$. A nonzero $t_s$ adds a phase term $\omega t_s$ to
(\ref{eq:d}). The total current can be obtained summing up all individual currents
given by (\ref{eq:c3}) with all relevant quantities labelled by $s$. 
Then, the total energy spectrum is derived as 
\begin{eqnarray}
U_{\rm tot}(\omega,\bhk)&\approx&
{e^2\omega^2\over16\pi^3\varepsilon_0c}\sum^N_{s,s'=1}\biggl[
\eta_s\eta_{s'}\,(\alpha_s-\theta)(\alpha_{s'}-\theta)F_sF^*_{s'}\nonumber\\
&&+
{\Omega^2_e\over\gamma_s\gamma_{s'}}\,
\sin\alpha_s\sin\alpha_{s'}\,F'_s{F'}^*_{s'}\biggr]\,e^{i\varphi_{ss'}},
 \label{eq:sp4}
\end{eqnarray}
where $N$ is the
total number of charged particles.
The coherence is determined by the phase $\varphi_{ss'}=d_s-d_{s'}$, given by
\begin{eqnarray}
\varphi_{ss'}&\approx&
\omega(t_s-t_{s'})-{\bk}\cdot({\bx}_{0s}-{\bx}_{0s'})\nonumber\\
&&-{\omega\over\Omega_s}
\left[1-n\beta_s\cos(\theta-\alpha_s)\right](\phi-\psi_{0s})
\nonumber\\
&& +{\omega\over\Omega_{s'}}
\left[1-n\beta_{s'}\cos(\theta-\alpha_{s'})\right](\phi-\psi_{0s'}).
\label{eq:psi2}
\end{eqnarray}
We use the following expression for the initial position:
 ${\bx}_0={\bx}_{0c}-(v_\perp/\Omega)(\sin\psi_0\,
{\be}_x-\cos\psi_0\,{\be}_y)$. The usual spontaneous emission 
corresponds to that the phase is $|\varphi_{ss'}|\gg1$ for $s\neq s'$
and that only terms of $s=s'$ contribute to the total spectrum. 
Therefore in the case  of spontaneous synchrotron radiation, 
the total spectrum can be written as $U_{\rm tot}=N\bar{U}$, where 
$\bar{U}$ is the single particle's spectrum
averaged over the particles' momentum distribution.
Since the initial gyrophase does not enter the final form of the energy spectrum,
spontaneous synchrotron radiation is axially symmetric with respect 
to the magnetic field line direction, i.e. the angular pattern of emission
depends only on $\theta$ not $\phi$. In the case of coherent synchrotron emission 
(cf. Sec. 3), such symmetry is broken since the maximum coherence 
depends explicitly on the initial gyrophases (cf. Eq. ~\ref{eq:psi2}).

\section{Coherent synchrotron emission}

Coherent emission occurs provided that the majority of emitting particles 
satisfy the condition $|\varphi_s-\varphi_{s'}|\ll1$. In general, one can calculate 
the total spectrum numerically using (\ref{eq:sp4}) 
for a given distribution of the particle injection time,
position and initial velocity. 
In some special cases, one may write down
its analytical form. The simplest case is where all 
particles have the same initial velocity, which is considered here. 

\begin{figure}
\includegraphics[width=7cm]{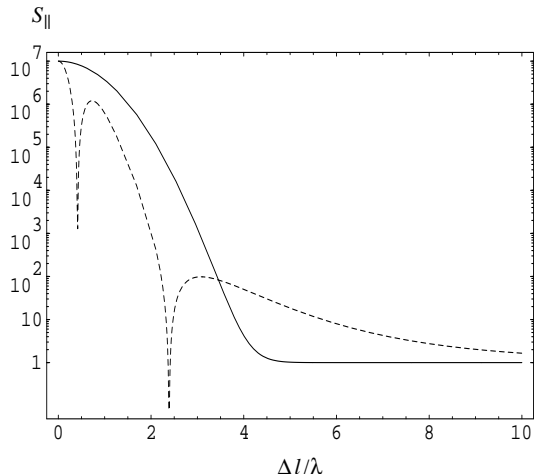}
\caption{Coherence factor vs bunch size. The solid and dashed
lines correspond respectively to a gaussian bunch (\ref{eq:Pgaussian}) and
a bunch of $\Gamma$-form (\ref{eq:gpdf}) with $A=5$.
One assumes the total number of emitting particles $N=10^7$. The bunch size is 
in units of the wavelength $\lambda$.}
\label{fig:coh}
\end{figure}

\begin{figure}
\includegraphics[width=7.5cm]{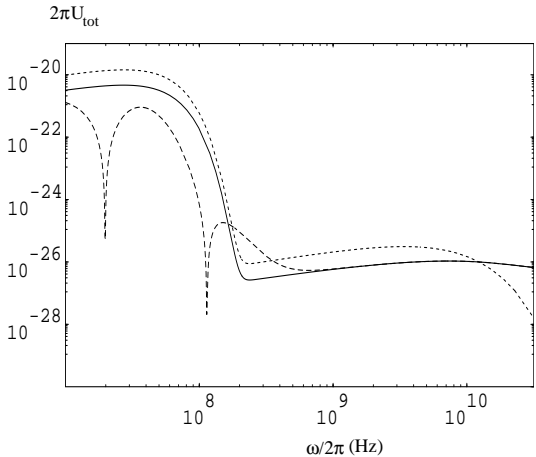}
\caption{Spectrum $2\pi U_{\rm tot}$ (in ${\rm J}\,{\rm Hz}^{-1}\,{\rm sr}^{-1}$)
from a gaussian bunch (solid line) with a longitudinal size $\Delta l=1\,\rm m$
and the $\Gamma$-pdf bunch (long-dashed line) with $A=3$.
The short-dashed line corresponds to the spectrum derived by ignoring the
prompt bremstrahlung. The parameters are as in figure~\ref{fig:spectrum1}.  
There is a critical frequency $\omega_{\rm coh}$ that separates the 
spectrum into the coherent region $\omega<\omega_{\rm coh}$ and
spontaneous emission region $\omega\geq\omega_{\rm coh}$.
For the gaussian bunch, the critical frequency is
$\omega_{\rm coh}/2\pi\sim (2\ln N)^{1/2}c/2\pi\Delta l\sim 270\,\rm MHz$ 
and for the $\Gamma$-pdf bunch it is $\omega_{\rm coh}/2\pi\sim
(c/\Delta l)N^{1/(1+A)}\approx 1\,\rm GHz$.
}
\label{fig:spectrumc1}
\end{figure}

One may write the total spectrum into the form
\begin{equation}
U_{\rm tot}=N\Bigl[S_\perp(\omega) U_\perp+S_\parallel(\omega)U_\parallel
\Bigr],
\label{eq:sp3}
\end{equation}
where $1\leq S_{\parallel,\perp}(\omega)\leq N$ is called the coherence factor. 
The value $S_{\parallel,\perp}(\omega)=N$
corresponds to completely coherent emission and $S_{\parallel,\perp}(\omega)=1$ to spontaneous emission.  
When the number densities of electrons and positrons are equal,
$j_{\perp\theta}$ does not contribute to the coherent power as 
contributions from electrons and positrons cancel out. This leads to $S_\perp=1$. The 
total spectrum is
\begin{eqnarray}
U_{\rm tot}&=&N\Bigl[U_\perp+S_\parallel(\omega)U_\parallel\Bigr], \\
U_\parallel&\approx&{e^2\over
16\pi\varepsilon_0c\sin\alpha}\left({4\omega\over\Omega_e\sin\alpha}\right)^{2/3}\nonumber\\
&&\times\biggl[{{\rm Ai}'}^2(\xi)+{{\rm Gi}'}^2(\xi)\biggr],
\label{eq:sp5}
\end{eqnarray}
where $\xi\approx (\omega/2\Omega\sin\alpha)^{2/3}[\gamma^{-2}+(\theta-\alpha)^2]$ and
$n=1$ is assumed. In contrast to coherent Cerenkov emission, which requires a charge asymmetry
(net charge) (Askar'yan 1962, 1965), coherent synchrotron emission can occur for a neutral plasma. 
In the case of charge symmetry, the polarization is linear, perpendicular to the
plane of the magnetic field and wave vector.
The coherence factor is then given by
\begin{equation}
S_\parallel(\omega)={1\over N}\sum^N_{s,s'=1}\cos\left[\omega(t_s-t_{s'})-{\bk}\cdot
({\bx}_{0s}-{\bx}_{0s'})\right].
\label{eq:Sp}
\end{equation}
Eq. (\ref{eq:Sp}) is sum of phasors, which can be modeled
as (e.g. Hartemann 2000) 
\begin{eqnarray}
S_\parallel(\omega)&\approx& 1-\langle\cos\Theta\rangle^2+N\langle\cos\Theta\rangle^2\nonumber\\
&&+2\langle\cos\Theta\rangle
\Bigl[N(1-\langle\cos\Theta\rangle^2)\Bigr]^{1/2},
\label{eq:Sp2}
\end{eqnarray} 
where $N\gg1$, $\Theta= \omega(t_s-{\bn}\cdot{\bx}_{0s}/c)$, ${\bn}={\bk}c/\omega$,
and the average is made over a distribution $P(\Theta)$,
\begin{equation}
\langle\cos\Theta\rangle\equiv\int^{2\pi}_0d\Theta\,P(\Theta)\cos\Theta.
\end{equation}
For a uniform distribution $P(\Theta)=1$, corresponding to spontaneous emission,
one has $S_\parallel(\omega)=1$, and for $P(\Theta)=\delta(\Theta)$, one has
$S_\parallel(\omega)=N$, corresponding to completely coherent emission.
As an example, we consider the case in which particles are injected at the 
same time, say at $t_s=0$, with a gaussian profile
in the longitudinal (along $\bk$) spatial distribution with
a width $\Delta l$. The probability can be written as
\begin{equation}
P(\Theta)={1\over\sqrt{\pi}\Delta\Theta}e^{-(\Theta/\Delta\Theta)^2},
\label{eq:Pgaussian}
\end{equation}
where $\Delta\Theta=\omega\Delta l/c$.
Then, one finds $\langle\cos\Theta\rangle=\exp(-\Delta\Theta^2/2)=
\exp[-(\omega\Delta l/c)^2/2]$. 
A plot of $S_\parallel$ as a function of $\Delta l$ is shown in figure~\ref{fig:coh}.
Similarly, if all particles are injected at the same 
location with a gaussian profile with a width $\Delta t$ in time, one 
has $\langle\cos\Theta\rangle=\exp[-(\omega\Delta t)^2/2]$. This shows that to
attain effective coherence, the spread in time of particle injection 
must be $\Delta t\sim \sqrt{2}/\omega$. The spectrum from a gaussian bunch is calculated using
(\ref{eq:sp3}) and is shown in figure~\ref{fig:spectrumc1}.  
The spectrum consists of two regions, separated 
by a critical frequency $\omega_{\rm coh}\sim (2\ln N)^{1/2}c/\Delta l$:
the coherent ($S_\parallel>1$) emission region $\omega<\omega_{\rm coh}$, where 
the polarization is predominantly linear, and spontaneous ($S_\parallel=1$)
emission region $\omega\geq\omega_{\rm coh}$, where the polarization
is elliptical as both $U_\parallel$ and $U_\perp$ are present.
The calculation of the coherence factor can easily
be extended to other types of distribution. In particular, 
the $\Gamma$-probability distribution is thought to be the more
relevant for air showers \citep{a-etal97}, which is discussed 
in detail in Sec. 4.

As the secondary particles from air showers have a distribution in momentum,
one needs to use the full expression (\ref{eq:sp4}) to calculate the spectrum. 
When the particle's momentum distribution is included, an analytical form for the spectrum 
can be derived only in some special cases. For example, when the bunch size is smaller than the 
relevant wavelength, one may ignore the phase term (\ref{eq:psi2}) and 
expresses the spectrum as a integration of the single particle formula
over the particle's distribution. The total intensity is about $N^2$ times 
larger than that by single particle.

\section{Application to air showers}

The formalism developed in the previous sections can be applied to
radio emission from extensive air showers (EAS). 
Detailed modeling requires 
numerical modeling of EAS, which was already considered by several authors  
\citep{sgr03,hf05a,hf05b}. Here we only discuss qualitatively  
implications of the modified synchrotron emission for radio emission from EAS
and further detailed modeling will be considered elsewhere. 
A notable feature of the modified synchrotron emission is that the corresponding
intensity derived from a semi-inifinite orbit is smaller than the 
conventional synchrotron emission by about a factor of 4. 
In principle, this feature can be tested against observations provided 
accurate radio spectra with well calibrated intensities become available,
for example with the planned Low Frequency Array (LOFAR)~\citep{fg03}.
Other features include an angular broadening of the radiation 
pattern and a slower decay of the spectrum above the critical frequency ($\omega\sim
\Omega_e\gamma^2$). The broadening due to prompt 
bremstrahlung is more pronounced for the perpendicular polarization than 
for the parallel polarization. Since the perpendicularly polarized
component depends on the charge excess, such broadening is important
when there is significant charge asymmetry. 
The spectral hardening occurs near the synchrotron
cut-off frequency, which is much higher than the transition
frequency $\omega_{\rm coh}$. So, this spectral feature may not be
observable for air showers.

In the following we derive the coherence factor using the procedure
in Sec. 3. As an example, one assumes that an air shower occurs at a few km height
and that all emitting particles are located in the shower maximum.
As the zeroth order approximation, the near field effect can be ignored and 
the energy spectrum (Eq. \ref{eq:sp4}) derived in Sec. 2 is applicable. 
The near field effect may need to be included if
the maximum of an air shower develops very near the ground level. There are extensive 
discussions of air showers in the literature \citep{g90} and in principle, one can obtain a 
distribution of particle's injection time ($t_s$), initial momentum ($\gamma{\bv}_{0s}$) 
and position (${\bx}_{0s}$). Assuming the primary cosmic ray energy to be
$E_p$, the number of secondary electrons and positrons can be estimated as 
$N\sim E_p/\gamma m_ec^2\sim 10^7$ for $E_p\sim 10^{15}\,\rm eV$ and $\gamma=
80$. In the practical situation these particles are injected over an extended range 
rather than at a single fixed height. For EAS, the distribution in $t_s$ for secondary 
electrons/positrons can be inferred from measurements of arrival times of charged particles 
(muons plus electrons/positrons).  One should note that these measured times are not actual 
arrival times of electrons and positrons
since muons arrive earlier than electrons/positrons.
The arrival times can be modeled by a $\Gamma$-probability distribution 
function ($\Gamma$-pdf)~\citep{a-etal97,a-etal01}.
Thus, the probability can be written as~\citep{hf03}:
\begin{equation}
P(\Theta)={1\over\Delta\Theta\Gamma(1+A)}\,\left({\Theta\over
\Delta\Theta}\right)^A\,e^{-\Theta/\Delta\Theta},
\label{eq:gpdf}
\end{equation}
for $\Theta>0$ and $P(\Theta)=0$ for $\Theta\leq0$, where
$\Gamma(x)$ is the Gamma function and the power index $A$ can be 
estimated from the distribution of electrons arrival times. The standard 
deviation is $\sigma_\Gamma=(1+A)^{1/2}\Delta\Theta$. 
From (\ref{eq:gpdf}), one obtains 
\begin{equation}
\langle\cos\Theta\rangle=\left(1+\Delta\Theta^2\right)^{-(1+A)/2}
\cos\biggl[(1+A)\arctan\Delta\Theta\biggr],
\label{eq:avcos}
\end{equation}
\begin{equation}
\langle \Theta\rangle= (1+A)\Delta\Theta.
\end{equation}
The corresponding critical frequency is given by $\omega_{\rm coh}\sim (c/\Delta l)N^{1/(1+A)}$. 
This frequency defines a transition from coherent to incoherent emission.
For $\Delta l=1\,\rm m$, $N=10^7$ and $A=5$, one has $\omega_{\rm coh}/2\pi\sim 1.2\,\rm GHz$. 
The coherence factor can be obtained by substituting (\ref{eq:avcos})
for (\ref{eq:Sp2}). Here we write down the two limiting cases:
\begin{equation}
S_\parallel\approx\left\{
\begin{array}{ll}
N(1+\Delta\Theta^2)^{-1-A} & \\
 \quad\quad\times\cos^2\biggl[
(1+A)\arctan\Delta\Theta\biggr], & \omega < \omega_{\rm coh},\\
1, & \omega \gg\omega_{\rm coh}.
\end{array}
\right.
\label{eq:Sp3}
\end{equation}
The dashed line in figure~\ref{fig:coh} shows $S_\parallel$ as a function of the bunch size 
$\Delta l$, featuring a higher transition frequency $\omega_{\rm coh}$
than the gaussian bunch. The low-frequency approximation in (\ref{eq:Sp3}) 
corresponds to the nonsquared form given by~\cite{hf03} except 
that the cosine factor is retained here.

The coherent emission can be elliptically polarized if there is a charge excess as it 
is the case for air showers. Assuming the excess number of electrons is $N_c$, one has 
\begin{equation}
{S_\perp\over S_\parallel}\sim {N_c\over N}.
\end{equation}
The right-hand side is sensitive to the cut-off energy of the excess electron energy. For 
a cut-off near the MeV energy, the excess of electrons can reach $N_c/N\sim 10-15\%$. 
(Note that cascades in a dense medium 
such as rocks can give rise to about $N_c/N\sim 20\%$ excess negative charge.) 
As a result, the coherent emission can be elliptically polarized with 
an ellipticity $\sim (S_\perp/S_\parallel)^{1/2}\delta_c
=(N_c/N)^2\delta_c \sim 0.3\delta_c$, where $\delta_c$ is the typical ellipticity
of single particle's radiation.

\section{Conclusions}

Synchrotron emission by a particle moving along a semi-infinite trajectory 
is considered. Since effective coherent synchrotron emission by secondary 
particles in an air shower occurs in the core of the shower where most emitting 
particles are created, the initial point of the particle's trajectory need be 
included explicitly. It is shown that radiation from a particle moving along a 
semi-infinite track can be separated into the usual synchrotron emission and the bremsstrahlung-like
emission (MIB). The latter is due to emission as the result of onset of
the particle's curved trajectory.  The spectral intensity of the modified synchrotron emission
(usual synchrotron plus MIB) is lower than that for normal synchrotron emission, roughly by a factor of 4.
It is interesting to note that such reduction is consistent with the recent 
result from numerical simulation by~\cite{hf05b}. The radiation pattern has a broader angular 
distribution than the usual synchrotron emission. This feature is 
especially pronounced for the perpendicularly
polarized component. The spectrum has a much slower decay above the 
critical frequency $\omega\sim\gamma^2\Omega_e$, while the usual
synchrotron spectrum has an exponential cutoff above
the critical frequency. In the application to radio emission from air showers,
the reduced intensity can be verified in principle provided 
accurate observations of the radio spectrum are available. Although there
exist early observations of air shower radio emission, there are some uncertainties
in determining the calibration factor \citep{a71,a78}. The current LOFAR Prototype 
Station (LOPES) and the future LOFAR may provide a better opportunity to test the 
predicted spectrum. Since the transition frequency (to spontaneous emission) 
is much lower than the cut-off frequency, change near the cut-off frequency 
may not be observable. The broadening of the radiation pattern
occurs mainly for the perpendicularly polarized component, which may be 
observable provided that there is significant charge asymmetry in the emitting
plasma. 

A major advantage of the formalism presented here is that 
the initial conditions including the time of particle creation, initial velocity
and gyrophase all appear in the phase (\ref{eq:psi2}) and these quantities can be 
modeled statistically. Although the single-particle formalism was used by~\cite{wm82}
to treat coherent gyromagnetic emission, in their calculation,
the finite track effect was not considered. It is shown here
that the distribution of the particle injection time is important in determining 
the coherence. The semi-infinite track approximation adopted here is  
valid provided that the emitting particles are highly relativistic with
$\gamma\gg1$. In the relativistic limit, the synchrotron pulse duration ($1/\Omega_e\gamma^2
\sim 10^{-8}\,\rm s$) is much shorter than the typical flight time $T\sim 10^{-6}\,\rm s$ 
for a particle's free path 500 m, and therefore the orbit can be 
approximately regarded as semi-infinite. If electrons and positrons from an air
shower have an energy cutoff extending to MeV energies with $\gamma\sim 1$, 
the synchrotron approximation is no longer valid and one has
cyclotron emission instead. In this case, a finite orbit needs to be considered.
One may extend the calculation in Sec. 3 and 4 to include the particle 
distribution in momentum and this requires
a numerical approach, which is not considered here.


\end{document}